\documentclass[12pt,a4paper]{article}
\usepackage{verbatim} 
\usepackage{amsmath} %
\usepackage{amssymb} %
\usepackage{theorem} %
\theorembodyfont{\itshape} \theoremstyle{plain}
\newtheorem{Theorem}{Theorem}[section]
\theorembodyfont{\itshape} \theoremstyle{plain}
\newtheorem{Corollary}{Corollary}[section]
\theorembodyfont{\itshape} \theoremstyle{plain}

\author{Pavel Exner$^{a,b}$, Helmut Linde$^{c,d}$, and Timo Weidl$^c$}
\title{Lieb-Thirring inequalities \\ for geometrically induced bound states}
\date{\small\itshape a) Nuclear Physics Institute, Academy of Sciences,
25068 \v{R}e\v{z} near Prague,Czechia
\\
b) Doppler Institute, Czech Technical University,
B\v{r}ehov{\'a}~7, 11519 Prague, Czechia
\\
c) Institute for Analysis, Dynamics and Modeling,
Faculty of Mathematics and Physics, Stuttgart University,
PF 80 11 40, D-70511 Stuttgart, Germany\\
d) Department of Physics,
Pontificia Universidad Catolica de Chile
Casilla 306, Correo 22 Santiago, Chile.}

\begin{document}
\maketitle

\begin{abstract}
\noindent We prove new inequalities of the Lieb-Thirring type on
the eigenvalues of Schr\"odinger operators in wave guides with
local perturbations. The estimates are optimal in the
weak-coupling case. To illustrate their applications, we consider,
in particular, a straight strip and a straight circular tube with
either mixed boundary conditions or boundary deformations.
\end{abstract}

\section{Introduction} \label{SectionIntroduction}

Recent progress in experimental physics provides various examples
of guided particles: electrons in semiconductor quantum wires or
carbon nanotubes, atoms in hollow fibers, etc. Moreover, there is
a close analogy between two-dimensional systems of this type and
flat microwave resonators -- see \cite{DE, Ku, LCM} for more
details and bibliography. The most simple model of such quantum
wave guides is a one-particle Schr\"odinger operator in a domain
of a strip or tube form subject to various boundary conditions. If
no external field is present, the stationary part of the problem,
in particular the search for bound states, is then reduced to
spectral analysis of the Laplace operator in such domains.

Consider the Dirichlet Laplacian on a straight tube
$\mathbb{R}\times\omega_0$ with a rather general cross-section
$\omega_0\subset \mathbb{R}^{d-1}$. The spectrum of this operator
is obviously purely absolutely continuous and it covers the
interval $[\lambda_1(\omega_0),\infty)$, where
$\lambda_1(\omega_0)$ is the lowest eigenvalue of the Dirichlet
Laplacian on $\omega_0$. If this ideal wave guide is perturbed,
for example, by local deformations or by a local change of the
boundary conditions, eigenvalues below the threshold $\lambda_0$
can appear. The corresponding bound states are sometimes called in
the literature trapped modes; the corresponding electron wave
functions are localized in the vicinity of the perturbation. This
effect is well studied and, in particular, the asymptotic behavior
of these eigenvalues for gentle deformations or small
perturbations of the boundary conditions has been investigated in
several papers, see e.g. \cite{BEG, BullaWeakly, DE, EK,
ExnerVugalterAsy, ExnerVugalterBound} and references therein.

On the other hand, only few quantitative results are known in the
non-asymptotic regime. Here one looks for estimates on the
discrete spectrum, such as the counting function \cite{ESTV,
ExnerVugalterBound} of the trapped modes or their Riesz means
\cite{ExnerWeidlLieb}. In the last named paper it has been shown
that due to the special geometry of mixed dimensionality of
quantum wave guides, operator-valued Lieb-Thirring inequalities
represent a suitable tool to tackle this problem. This was then
applied to a straight wave guide with an attractive potential
interaction. In the present work we are going to demonstrate how a
similar approach can yield estimates for the case of locally
deformed ``quantum wires'' or for bound states induced by a local
modification of boundary conditions.

\section{Preliminary about Lieb-Thirring inequalities} \label{AuxMatLTh}

The aim of this section is to collect an auxiliary material on
Lieb-Thirring estimates on \( L^{2}(\mathbb{R}^{d}) \), which
shall be of use in the following.

Let \( \mathcal{G} \) be a separable Hilbert space and let \( W \)
be a function on \( {\Bbb R}^{d} \) which takes almost everywhere
non-negative compact operators on \( \mathcal{G} \) as its values.
We consider eigenvalue moments of the Schr{\"o}dinger type
operator
\[
H=1_{\mathcal{G}}\otimes (-\Delta )-W(x)\quad \mbox {on}\quad
\mathcal{G}\otimes L^{2}({\Bbb R}^{d})\,.\]
Suppose that \( tr_{\mathcal{G}}W^{\sigma+\frac{d}{2}}(\cdot )\in
L^{\sigma +\frac{d}{2}}({\Bbb R}^{d}) \). Then for \( \sigma \geq
1/2 \) if \( d=1 \), and for \( \sigma >0 \) if \( d\geq 2 \), the
following estimate holds true\footnote{Throughout the paper, we
use the notation $x_\pm := (|x|\pm x)/2$ for the positive and
negative part of numbers, functions or self-adjoint operators,
respectively.}:
\begin{equation}
\label{es12} tr_{\mathcal{G}\times L^{2}({\Bbb R}^{d})}\,
H_{-}^{\sigma }\leq r(\sigma ,d)L_{\sigma ,d}^{cl}\int _{{\Bbb
R}^{d}}tr_{\mathcal{G}}\, W^{\sigma +\frac{d}{2}}(x)
\mathrm{d}x\,,
\end{equation}
where
\[
L_{\sigma ,d}^{cl}:=\frac{\Gamma (\sigma +1)}{2^{d}\pi
^{d/2}\Gamma (\sigma +\frac{d}{2}+1)}\,.\]
Moreover, the constants \( r(\sigma ,d) \) in (\ref{es12}) satisfy
the inequalities
\begin{eqnarray}
r(\sigma ,d)=1 & \quad \mbox {if}\quad  & \sigma \geq 3/2,\, d\in {\Bbb N},\label{R1} \\
r(\sigma ,d)\leq 2 & \quad \mbox {if}\quad  & 1\leq \sigma <3/2,\, d\in {\Bbb N},\label{R2} \\
r(\sigma ,d)\leq 2 & \quad \mbox {if}\quad  & 1/2\leq \sigma <1,\, d=1,\label{R3} \\
r(\sigma ,d)\leq 4 & \quad \mbox {if}\quad  & 1/2\leq \sigma <1,\, d\in {\Bbb N},\, d\geq 2,\label{R4}
\end{eqnarray}
see \cite{LW,HLW,HLT,Hundertmark}. Usually these inequalities are
stated for the scalar operator
\[
H_{\alpha }=-\Delta -\alpha V\quad \mbox {on}\quad L^{2}({\Bbb
R}^{d})\,,\]
i.e. for $\mathcal{G}=\mathbb{C}$, see \cite{LT,C,L,R} and
\cite{W,HLT}; then these bounds give estimates on spectral
quantities in terms of the classical phase space volume.

The generalization \eqref{es12} to operator-valued potentials has
been the crucial step for the recent progress on the constants in
Lieb-Thirring inequalities in higher dimensions. The idea of
``lifting'' in dimensions, given in \cite{LW}, is also the base
for the proof of the main result of this paper.

\section{Statement of the result} \label{SectionResult}

Consider an open set $\Omega \subset \mathbb R^{d}$. Let
$(x_1,\dots, x_d)$ be the Cartesian coordinates in $\mathbb
R^{d}$. For a vector $x\in\mathbb R^{d}$ we shall single out the
first coordinate and write $x=(\xi,\eta)$ with
$\eta=(x_2,\dots,x_{d})\in\mathbb R^{d-1}$ and
$\xi=x_1\in\mathbb{R}$. For a given value of $\xi$ let
\begin{equation*}
\omega(\xi)=\{\eta\in\mathbb R^{d-1}|\; x=(\xi,\eta)\in \Omega\,\}
\end{equation*}
be the cross-section of $\Omega$ at the point $\xi$ which is an
open set in $\mathbb R^{d-1}$.  We shall assume that the sets
$\omega(\xi)$ are uniformly bounded and non-empty for any
$\xi\in\mathbb R$, and that $\Omega$ is a straight tube with local
perturbations, that is
\begin{equation*}\label{omega0}
\omega(\xi)=\omega_0\quad\mbox{for all}\quad |\xi|>R\, .
\end{equation*}
for some open set $\omega_0$ and a positive $R$. The local
deformation of $\Omega$ is given by the shape of the
cross-sections $\omega(\xi)$.

Consider further a set $\Gamma\subset\overline{\Omega}$, such that
$\Omega\setminus \Gamma$ is open and that its projection onto the
transverse plane,
\begin{equation*}
P_\Gamma:=\{\eta\in\mathbb R^{d-1}|\, \exists \xi\in\mathbb{R}
\;\:\mathrm{such\;that}\; x=(\xi,\eta)\in \Gamma\}\,,
\end{equation*}
has zero Lebesgue measure in $\mathbb R^{d-1}$.

Let $-\Delta^\Omega_\Gamma$ be the self-adjoint realization of the
Laplace operator on $L^2(\Omega\setminus\Gamma)$ with Dirichlet
conditions on $\partial\Omega\setminus\Gamma$ and Neumann
conditions on $\Gamma$. This means that the quadratic form
$$\int_{\Omega\setminus\Gamma}|\nabla u|^2\mathrm{d}^dx$$
generating the operator $-\Delta^\Omega_\Gamma$ is defined on the
closure (with respect to the $W^{1,2}$ Sobolev norm) of the set of
all smooth functions in $\Omega\setminus\Gamma$, which vanish for
large $|\xi|$ and in a vicinity of $\partial\Omega\setminus
\Gamma$ and which are square integrable together with their first
partial derivatives. For a fixed $\xi\in\mathbb R$ we define
\begin{equation*}
\gamma(\xi)=\{\eta\in\mathbb R^{d-1}|\; x=(\xi,\eta)\in\Gamma\}\,.
\end{equation*}
As above let $-\Delta^{\omega}_{\gamma}$ be the self-adjoint
realization of the Laplace operator on $\omega\setminus\gamma$
with Dirichlet conditions on $\partial\omega\setminus\gamma$ and
Neumann conditions on $\gamma$, where $\omega=\omega(\xi)$ and
$\gamma=\gamma(\xi)$. Under suitable conditions on $\gamma$ the
spectrum (or at least the lower portion of it) is
discrete\footnote{In general, this is the case unless the set
$\gamma$ is too ``wild'' -- see, e.g., \cite{HSS, Si2}.}. In this
case the corresponding eigenvalues will be denoted by
$\lambda_j(\omega,\gamma)\,,\; j=1,2,\dots\,$; if
$\gamma=\emptyset$ we shall simply write $\lambda_j(\omega)$
instead of $\lambda_j(\omega,\emptyset)$. Of particular importance
is the ``asymptotic'' quantity $\lambda_1(\omega_0)$ with
$\omega_0$ from eq.~\eqref{omega0}. We will suppose that the
functions $\xi\mapsto \lambda_j(\omega(\xi),\gamma(\xi))$ are
measurable\footnote{ This requirement imposes again a restriction
on the geometry of $\Omega$ and $\Gamma$. For instance, in the
pure Dirichlet case, $\Gamma= \emptyset$, this property is
guaranteed provided that, apart of a discrete subset of $[-R,R]$,
to each $\xi$ and $\varepsilon>0$ there is an open set $O\ni \xi$
such that for any $\xi'\in O$ the symmetric difference
$\omega(\xi)\Delta \omega(\xi')$ is contained in the
$\varepsilon$-neighborhood of the boundary $\partial\omega(\xi)$,
because the eigenvalues are in this case piecewise continuous as
functions of $\xi$ -- cf.~\cite{RT}.}, or at least that this
property is valid below $\lambda_1(\omega_0)$.

Assume now that the spectrum of $-\Delta^\Omega_\Gamma$ below
$\lambda_1(\omega_0)$ is discrete. In general it may be empty, of
course; we are interested in situations when it is not. Then the
corresponding eigenvalues will be called $\Lambda_j(\Omega,\Gamma)
\,,\; j=1,2,\dots\,,$ and in case of $\Gamma=\emptyset$ we write
$\Lambda_j(\Omega)$ instead of $\Lambda_j(\Omega,\emptyset)$. In
particular, if there is only one such eigenvalue we drop the index
$j$. It is convenient to define the ``shifted'' operator
\begin{equation*}
H := -\Delta^\Omega_\Gamma - \lambda_1(\omega_0)
\end{equation*}
on $L^2(\Omega\setminus\Gamma)$, the essential spectrum of which
is by assumption and an elementary bracketing argument equal to
\begin{equation*}
\sigma_\mathrm{ess}(H) = [0,\infty)\,,
\end{equation*}
while the perturbation can give rise to bound states of negative
energy. The following estimate on the moments of these negative
eigenvalues is the main result of this paper:

\begin{Theorem}
\label{TheoremGeneral} Suppose that the spectrum of the operators
$-\Delta^\omega_\gamma$ with $\omega=\omega(\xi)$ and
$\gamma=\gamma(\xi)$ below $\lambda_1(\omega_0)$ is discrete and
finite for almost all $\xi\in\mathbb R$, the eigenvalues are
measurable w.r.t. $\xi$, and that
\begin{equation*}
I_{\Omega,\Gamma,\sigma}:=\int_{\mathbb R}\mathrm{tr } \left(
-\Delta^{\omega(\xi)}_{\gamma(\xi)} - \lambda_1(\omega_0)
\right)_-^{\sigma+1/2} \mathrm{d}\xi = \int_{\mathbb R} \sum_j
\big( \lambda_j(\omega(\xi),\gamma(\xi)) - \lambda_1(\omega_0)
\big)_-^{\sigma+1/2} \mathrm{d}\xi
\end{equation*}
is finite for $\sigma \ge 1/2$. Then the negative spectrum of $H$
is discrete and the inequality
\begin{equation} \label{EquationTheorem}
\mathrm{tr}\, H_-^\sigma \le r(\sigma,1)\,L^\mathrm{cl}_{\sigma,1}
I_{\Omega,\Gamma,\sigma}
\end{equation}
holds true.
\end{Theorem}

We will prove Theorem \ref{TheoremGeneral} in
Sec.~\ref{SectionProofGeneral}. Before doing that we notice that
it applies to a variety of particular cases, a selection of which
is given in the following section.

\section{Examples} \label{SectionApplication}

\subsection{Strip with a Neumann perturbation} \label{SubsectionStripNeumann}

Let $\Omega = \mathbb R \times (0,1)$ be a planar strip and
$\Gamma = [0,\alpha] \times \{b\}$ with $\alpha>0$ and $\frac 12 <
b \le 1$ a line segment in the interior or on the boundary of
$\Omega$, away of the strip axis. Then the cross-section of the
strip is $\omega(\xi) = \omega_0 = (0,1)$ while the cross-section
of $\Gamma$ is $\gamma(\xi) = \{b\}$ for $\xi \in [0,\alpha]$ and
$\gamma(\xi) = \emptyset$ otherwise. The spectrum of the Laplacian
$-\Delta^{\omega(\xi)}_{\gamma(\xi)}$ can be determined easily as
its eigenfunctions are simple sine functions. The lowest
eigenvalue of $-\Delta^{\omega_0}_{\emptyset}$ is
$\lambda_1(\omega_0) = \pi^2$, which is therefore also the lower
edge of the essential spectrum of $-\Delta^{\Omega}_{\Gamma}$. For
$\xi \in [0,\alpha]$, the operator
$\Delta^{\omega(\xi)}_{\gamma(\xi)}$ has a single eigenvalue
$\frac{\pi^2}{4b^2}$ below $\pi^2$. Combining this information
with (\ref{EquationTheorem}) we obtain:

\begin{Corollary}
\label{CorollaryStripNeumann} For $H = -\Delta^\Omega_\Gamma -
\pi^2$ and $\sigma \ge 1/2$ the following inequality is valid,
\begin{equation} \label{EquationCorollaryNeumann}
\mathrm{tr}\,H_-^\sigma \le
r(\sigma,1)\,L^\mathrm{cl}_{\sigma,1}\,\alpha
\left(\pi^2-\frac{\pi^2}{4b^2}\right)^{\sigma+1/2}.
\end{equation}
\end{Corollary}
The result remains valid, of course, for $b=\frac 12$ when it
becomes trivial.

\subsection{Strip with bulges}

Suppose now that $\Gamma = \emptyset$ and $\Omega_f = \{(\xi,\eta)
\in \mathbb R^2|\; 0<\eta<1+f(\xi)\}$ with a piecewise continuous
and compactly supported function $f$ such that $0\le f(\xi) < 1$.
Then we get in a similar way as above the following bound:

\begin{Corollary}
\label{CorollaryStripBumps} For $H = -\Delta^{\Omega_f} - \pi^2$
and $\sigma \ge 1/2$ we have
\begin{equation*}
\mathrm{tr}\,H_-^\sigma \le
r(\sigma,1)\,L^\mathrm{cl}_{\sigma,1}\, \pi^{2\sigma+1}
\int\limits_{-\infty}^\infty
\left(1-\frac{1}{(1+f(\xi))^2}\right)^{\sigma+1/2}
\mathrm{d}\xi\,.
\end{equation*}
\end{Corollary}
Note that the assumption $f(\xi) < 1$ is made here only for
simplicity; it ensures that $-\Delta^{\omega(\xi)}$ has not more
that one eigenvalue below $\pi^2$. It is straightforward to
generalize the claim to a more general profile function replacing
the integrand by $\sum_{j=1}^\infty \big( 1- j^2(1+f(\xi))^{-2}
\big)_{+}^{\sigma+1/2}$, where the sum has, of course, only a
finite number of nonzero terms for any fixed $\xi$.

\subsection{Circular tube with bulges} \label{defcirc}

As another particular case let us consider a tube in
$\mathbb{R}^3$ with Dirichlet boundary which is circular outside a
compact and has local bulges. The spectrum of the Laplace operator
on a circular disk with unit radius is well known: it is purely
discrete and expressed in terms of Bessel function zeros, in
particular, the lowest eigenvalue is $j_{0,1}^2$, where $j_{0,1}$
is the first positive root of the function $J_0$. It is also known
that among all domains of the same area, the first eigenvalue is
minimized by the circular disk; this fact is expressed in the
well-known Rayleigh-Faber-Krahn inequality \cite{Faber,Krahn}
\begin{equation} \label{EquFaberKrahn}
\lambda_1(\omega) \ge \frac{\pi j_{0,1}^2}{A(\omega)}\,,
\end{equation}
where $A(\omega)$ is the area of the domain $\omega$ and
$\lambda_1(\omega)= \inf\sigma\big(-\Delta^\omega\big)$.

We are again interested primarily in the situation when the bulge
is not too big. Notice that the second eigenvalue
$\lambda_2(\omega)$ can also be estimated with the help of
(\ref{EquFaberKrahn}): since $-\Delta^\omega$ commutes with the
involution defined by complex conjugation, the eigenfunction
$\Psi_2^{\omega}$ corresponding to $\lambda_2(\omega)$ can be
chosen as real-valued; it vanishes on a smooth nodal line without
endpoints in (the interior of) the cross section\footnote{The
shape of this nodal line depends on the cross section geometry. If
$\omega$ is simply connected the endpoints lie at the boundary,
while for a non-simply connected $\omega$ it may be also a closed
loop which does not touch the boundary \cite{HHON, Fo}.}. It
follows that this curve divides $\omega$ into two parts, one of
which must cover an area not exceeding $A(\omega)/2$; we call this
part $\tilde\omega$. Then $\Psi_2(\omega)$ restricted to
$\tilde\omega$ is also the ground-state eigenfunction of the
Dirichlet Laplacian $-\Delta^{\tilde\omega}$, thus
eq.~(\ref{EquFaberKrahn}) yields\footnote{The conclusion is not
affected by the fact that $\omega\setminus\tilde\omega$ may have a
larger area because the ground-state eigenvalue in the two parts
must be the same, of course.}
\begin{equation*}
\lambda_2(\omega) \ge \frac{\pi j_{0,1}^2}{A(\tilde\omega)} \ge
\frac{2\pi j_{0,1}^2}{A(\omega)}\,.
\end{equation*}
Consequently\footnote{In particular cases one can do better. For
instance, if the bulged tube is circular again, being described by
a radius function $r$, then there is a single transverse
eigenvalue below the threshold as long as $r(\xi)\le
j_{1,1}/j_{0,1}\approx 1.5933$ which means $A(\omega)\lesssim
2.5387\,\pi$.}, if $A(\omega) \le 2\pi$ then $\lambda_2(\omega)\ge
j_{0,1}^2$ so that $\lambda_1(\omega)$ is the only eigenvalue
which could be below $j_{0,1}^2$. It is indeed the case in the
bulged part of the tube where $\omega\setminus\omega_0$ has a
nonzero measure as it follows from the domain monotonicity of
Dirichlet eigenvalues \cite{GZ}. From the above remarks and
Theorem \ref{TheoremGeneral} we make the following conclusion:

\begin{Corollary}
\label{CorollaryTubeBumps} Define $\Omega$ as in
Sec.~\ref{SectionResult} with $\omega_0$ being a circular disk of
unit radius and $\Gamma = \emptyset$ and $\omega(\xi) \supset
\omega_0$ for all $\xi\in\mathbb{R}$. Moreover, suppose that the
area $A(\omega(\xi))$ of $\omega(\xi)$ satisfies $A(\omega(\xi))
\le 2\pi$. Then for $H = -\Delta^{\Omega} - j_{0,1}^2$ and $\sigma
\ge 1/2$ the inequality
\begin{equation*}
\mathrm{tr}\,H_-^\sigma \le
r(\sigma,1)\,L_{\sigma,1}^{cl}\,j_{0,1}^{2\sigma+1}
\int\limits_{-\infty}^\infty \mathrm{d}\xi\, \left(
1-\frac{\pi}{A(\omega(\xi))} \right)^{\sigma+1/2}
\end{equation*}
holds true.
\end{Corollary}

\section{Discussion of the results} \label{SectionDiscussion}

Let us next compare the obtained results with those of earlier
publications.

\subsection{Strip with a small bulge}

Consider the set $\Omega_{\alpha f}$ defined as in Corollary
\ref{CorollaryStripBumps} with the function $f$ replaced by
$\alpha f$ to have a parameter which controls the deformation. For
the sake of brevity we denote
\begin{equation*}
F_n:=\int_{-\infty}^\infty f(x)^n\,\mathrm{d}x\,.
\end{equation*}
It is known from \cite{BullaWeakly} that for a sufficiently smooth
$f$ and small $\alpha$ the operator $-\Delta^{\Omega_{\alpha f}}$
has exactly one eigenvalue below $\pi^2$ and its asymptotic
behavior is
\begin{equation*}
\Lambda(\Omega_{\alpha f}) = \pi^2 - \pi^4 F_1^2 \alpha^2 +
\mathcal{O}(\alpha^3)\,.
\end{equation*}
Expanding the estimate of Corollary \ref{CorollaryStripBumps} into
powers of $\alpha$, substituting $\pi^2 - \Lambda(\Omega_{\alpha
f})$ for $\textmd{tr }H_-$ and choosing $\sigma=\frac 12$, we
obtain
\begin{equation*}
\Lambda(\Omega_{\alpha f}) \ge \pi^2 - \pi^4 F_1^2\, \alpha^2 +
3\pi^4 F_1\, F_2 \,\alpha^3 -
\left(\frac{9}{4}F_2^2+4F_1F_3\right) \pi^4 \alpha^4 + \mathcal{
O}(\alpha^5)\,,
\end{equation*}
which means that our Lieb-Thirring inequality reproduces the true
weak-coupling asymptotics in this case.

\subsection{Strip with Neumann perturbation on the boundary}

The last claim need not be valid in general. Consider the set
$\Omega$ of Corollary \ref{CorollaryStripNeumann} with the
perturbation at the boundary, i.e. take $\Gamma_\alpha =
[0,\alpha]\times\{1\}$ with some $\alpha>0$. Then by \cite{ESTV}
the operator $-\Delta^\Omega_{\Gamma_\alpha}$ has for small enough
$\alpha$ exactly one eigenvalue below $\pi^2$. Choosing
$\sigma=\frac 12$, Corollary~\ref{CorollaryStripNeumann} yields
\begin{equation*}
\Lambda(\Omega,\Gamma_\alpha) \ge \pi^2 - \frac{9}{16} \pi^4 \alpha^2.
\end{equation*}
On the other hand it is known from \cite{ExnerVugalterAsy} that
for small $\alpha$ there are positive $c_1,c_2$ such
that\footnote{In fact, the eigenvalue has a Taylor expansion in
$\alpha$ and the coefficient of the leading fourth-order term can
be computed explicitly -- see \cite{Po} and also \cite{BEG}.}
\begin{equation*}
\pi^2 - c_1 \alpha^4 \le \Lambda(\Omega,\Gamma_\alpha)
\le \pi^2 - c_2 \alpha^4
\end{equation*}
holds, and consequently, our Lieb-Thirring inequality gives a too
rough weak-coupling estimate in this case.

On the other hand, the estimate is of a correct order in $\alpha$
in the strong coupling case, i.e. for large $\alpha$. To justify
this claim, recall a simple bracketing bound used in \cite{ESTV}.
The spectrum is estimated from above by means of adding extra
Dirichlet conditions at $\xi=0,a$ which yield the following
orthogonal family of functions,
\begin{equation*}
\Psi_n(\xi,\eta) := \left\{ \begin{array}{ll}%
\cos \frac{\pi}{2}\eta \,\sin \frac{n\pi}{\alpha} \xi &
\quad \mathrm{for}\; \xi \in [0,\alpha]\,,\\
0 &  \quad \mathrm{for}\;\xi \notin [0,\alpha]\,.\\
\end{array}\right.
\end{equation*}
This leads to a lower bound on $\mathrm{tr}\,H_-^\sigma$, namely
\begin{eqnarray*}
\rm{tr}\,H_-^\sigma &\ge& \sum\limits_{n=1}^{\infty}
\left(\frac{\pi^2}{4}+\frac{n^2\pi^2}{\alpha^2} -
\pi^2\right)_-^\sigma \\
&=& \pi^{2\sigma} \left(\frac {3}{4} \right)^{\sigma+1/2} \alpha
\int\limits_{0}^\infty (s^2 - 1)_-^\sigma \, \mathrm{d}s +
o(\alpha)\\
&=& L^\mathrm{cl}_{\sigma,1} \alpha \left(\frac {3
\pi^2}{4}\right)^{\sigma+1/2} + o(\alpha)\,.
\end{eqnarray*}
In a similar way Neumann bracketing provides an upper bound on
$\mathrm{tr}\,H_-^\sigma$ which differs from the lower one only by
the summation range which now starts from $n=0$, and hence gives
the same expression up to the error term. A comparison with
eq.~(\ref{EquationCorollaryNeumann}) for $b = 1$ shows that our
estimate exhibits the correct power of $\alpha$, the only
difference being the factor $r(\sigma,1)$ -- cf.~the relations
(\ref{R1})--(\ref{R4}).

\subsection{General considerations}

In the paper \cite{ExnerWeidlLieb} a similar formula has been
derived to estimate the moments of the binding energies in a
straight wave guide with an attractive potential. The estimating
expression differs from the r.h.s. of eq.~(\ref{EquationTheorem}):
it consists of two terms reflecting the mixed dimensionality of
the problem. One term describes the effect of a weak potential
where the dominating behavior of the eigenfunctions is
one-dimensional. The second one is important in the case of a
strongly attractive potential where the influence of the boundary
and the ``leads'' on the wave functions of the trapped particle in
the lower part of the spectrum is negligible and the problem is
essentially $d$-dimensional.

In the present work we have worked out estimates consisting of one
term only, having on mind in the first place systems which have no
more than one transverse eigenvalue below the threshold
$\lambda_1(\omega_0)$. This can still yield a good estimate if the
perturbation is rather ``long'' than ``wide'' as the previous
example illustrates. Moreover, spectra of wave guides with large
deformations can be well estimated by combination of bracketing
and standard phase-space methods.

Our result exhibits the usual Lieb-Thirring features in the sense
that it neglects repulsive components of the interaction, and the
bound may become useless if the latter dominate. Consider, for
instance, a deformed circular tube of Sec.~\ref{defcirc} and
suppose that the deformation is both squeezing and expanding the
cross section. If the cross section in the deformed part deviates
substantially from the circular shape, it may happen that the
discrete spectrum is empty even if the deformation adds volume to
$\Omega$ and the r.h.s. of the inequality in
Corollary~\ref{CorollaryTubeBumps} is nonzero.

For sake of simplicity we have limited our considerations to wave
guides which differ from a straight tube on a compact only. Some
generalizations would not be difficult to derive. For example, the
basic estimate (\ref{EquationTheorem}) of
Theorem~\ref{TheoremGeneral} will also hold true for a wave guide
the straight parts of which on both sides of the local
perturbation are parallel but not in line with each other. In a
similar way it is possible to generalize
Theorem~\ref{TheoremGeneral} to certain perturbations that are not
compactly supported but still local in the sense that they fall
off asymptotically fast enough. On the other hand, for instance,
it is not possible to extend our results in a straightforward
manner to the case of Neumann boundary conditions on a surface
which is not parallel to the tube axis; the reason will become
clear from the proof of Theorem \ref{TheoremGeneral} which we are
now finally going to present.

\section{Proof of Theorem \ref{TheoremGeneral}} \label{SectionProofGeneral}

As usual the (shifted) Laplace operator on $L^2(\Omega)$ is
associated with the closed quadratic form
\begin{equation}\label{hPsi}
h[\Psi,\Psi] = \int_\Omega \left(|\nabla \Psi|^2 -
\lambda_1(\omega_0) |\Psi|^2\right)\,\mathrm{d}x \,,
\end{equation}
where the boundary conditions are implemented by a proper choice
of the domain $Q(h)$ of the form $h$. In our case, when we deal
with $H=-\Delta^\Omega_\Gamma-\lambda_1(\omega_0)$, the form
domain $Q(h)$ is given by the $|\cdot|_h$-closure\footnote{Here
and in the following we use the symbol $|\cdot|_h$ for the
slightly modified Sobolev norm defined by $|\cdot|_h^2 =
h[\cdot,\cdot] + (\lambda_1(\omega_0) + 1) ||\cdot||^2$.} of the
set $M(\Omega,\Gamma)$ of all functions $\Psi\in
C^\infty(\Omega\setminus\Gamma)$, which vanish in the vicinity of
$\partial \Omega \setminus \Gamma$ as well as for sufficiently
large $|\xi|$, and for which the expression \eqref{hPsi} is
finite\footnote{In particular, such functions can have a ``jump''
on $\Gamma\cap\Omega$.}.

Now we define the smallest common envelope of the cross sections,
which is bounded by assumption, and the corresponding cylindrical
envelope of the tube by
\begin{equation*}
\hat\omega := \underset{\xi \in \mathbb R}{\bigcup}\: \omega(\xi)
\quad \textmd{and} \quad \hat\Omega := \mathbb R \times
\hat\omega\,,
\end{equation*}
so we have $\Omega \subset \hat\Omega$. Consider the quadratic
form on $L^2(\hat\Omega)$ given by
\begin{equation} \label{EquDefinitionHath}
\hat h[\Psi,\Psi] := \int\limits_\Omega \left(|\nabla \Psi|^2 -
\lambda_1(\omega_0) |\Psi|^2\right)\,\mathrm{d}x +
\int\limits_{\hat\Omega \setminus \Omega} \left|\frac{\partial
\Psi}{\partial \xi}\right|^2\,\mathrm{d}x
\end{equation}
with the form  domain $Q(\hat h)$ equal to the $|\cdot|_{\hat
h}$-closure of the set $\hat M(\Omega,\Gamma)$ of all functions
$\Psi \in L^2(\hat\Omega)$ for which $\Psi|_\Omega \in
M(\Omega,\Gamma)$ holds and the restriction $\Psi|_{\hat
\Omega\setminus\Omega}$ is smooth and vanishes near $\partial
\Omega$ and $\partial \hat\Omega$. Then $Q(\hat
h)=Q(h)\oplus_{\hat h} Y$ where the set $Y\subset
L^2(\hat\Omega\setminus\Omega)$ consists of all functions $\phi$
which are differentiable in the sense if distributions in the
$\xi$-direction and satisfy $\frac{\partial \phi}{\partial \xi}\in
L^2(\hat\Omega\setminus\Omega)$.

The closed quadratic form $\hat h$ is associated with the
self-adjoint operator
$$\hat H=H\oplus\left(-\frac{\partial^2}{\partial \xi^2}\right)
\quad\mbox{on}\quad L^2(\hat\Omega)=L^2(\Omega)
\oplus L^2(\hat\Omega\setminus\Omega),$$
which is the direct sum of our original operator $H$ on
$L^2(\Omega)$ and the differential operator
$-\frac{\partial^2}{\partial \xi^2}$ on
$L^2(\hat\Omega\setminus\Omega)$ with Dirichlet condition on the
part of $\partial\Omega$ which is not parallel to $\partial
\hat\Omega$. The last named operator is positive by definition,
and therefore $\hat H$ and $H$ have the same negative spectrum.

We can write the form $\hat h$ as a sum of parallel and transverse
components,
\begin{equation*}
\hat h[\Psi,\Psi] = \int_{\mathbb R}
\mathrm{d}\xi\,\left(\int_{\hat\omega} \mathrm{d}\eta\,
\left|\frac{\partial \Psi}{\partial \xi}\right|^2 +
w(\xi)[\Psi(\xi,\cdot),\Psi(\xi,\cdot)] \right),
\end{equation*}
with the second term defined through the quadratic form
\begin{equation*}
w(\xi)[\phi,\phi] := \int_{\omega(\xi)} \mathrm{d}\eta
\left[\left|\nabla_\eta\phi(\eta)\right|^2 - \lambda_1(\omega_0)
|\phi(\eta)|^2\right]\,.
\end{equation*}
The domain of $w(\xi)$ can be chosen as
\begin{equation*}
Q(w(\xi)) := \{\phi \in L^2(\hat\omega):
\phi|_{\omega(\xi)} \in Q(h(\xi))\},
\end{equation*}
where $Q(h(\xi))$ is the domain of the quadratic form
$h(\xi)[\phi,\phi]=\int_{\omega(\xi)}|\nabla_\eta \phi|^2
\mathrm{d}\eta$ associated with
$-\Delta^{\omega(\xi)}_{\gamma(\xi)}$ on $L^2(\omega(\xi))$.
Indeed, with such a domain choice we have $\Psi(\xi,\cdot)\in
Q(w(\xi))$ for any $\Psi \in Q(\hat h)$ and almost every $\xi \in
\mathbb R$.

It is straightforward to check that the form $w(\xi)$ is closed
and associated with the operator
\begin{equation*}
W(\xi) = \left[-\Delta^{\omega(\xi)}_{\gamma(\xi)} -
\lambda_1(\omega_0)\right] \oplus
0_{\hat\omega\setminus\omega(\xi)}\,,
\end{equation*}
where $0_{\hat\omega\setminus\omega(\xi)}$ is, of course, the zero
operator on $L^2(\hat\omega\setminus\omega(\xi))$. It follows from
the assumptions of Theorem~\ref{TheoremGeneral} that the negative
spectrum of $W(\xi)$ consists of at most finitely many negative
eigenvalues. Let $W_-(\xi)$ be the negative part of the operator
$W(\xi)$. Then $W_-(\xi)$ is an operator of finite rank on
$L^2(\hat\omega)$, and consequently, its  quadratic form
$w_-(\xi)$ is defined on $Q(w_-(\xi)) = L^2(\hat\omega)$.

Next we introduce the quadratic form
\begin{equation*}
\tilde h[\Psi,\Psi] := \int_{\mathbb R}
\left(\int_{\hat\omega}\left|\frac{\partial \Psi}{\partial
\xi}\right|^2\,\mathrm{d}\eta -
w_-(\xi)[\Psi(\xi,\cdot),\Psi(\xi,\cdot)] \right)d\xi\, ,
\end{equation*}
defined on the $|\cdot|_{\tilde h}$-closure of the set of all
smooth functions in $L^2(\hat\Omega)$. Making the closure
explicit, we find that $Q(\tilde h)$ consists of all functions
$\Psi\in L^2(\hat{\Omega})$ for which the following conditions
hold true: \vspace{-.8ex}
\begin{enumerate}
\item[(a)] for a.e. $\eta\in\hat{\omega}$ the function
$\Psi(\cdot,\eta)$ is differentiable in the sense of distributions
in $\xi$-direction on $\mathbb{R}$ and $\frac{\partial
\Psi}{\partial \xi}\in L^2(\hat{\Omega})$, \vspace{-.8ex}
\item[(b)]  for a.e. $\eta\in\hat{\omega}$ the function
$\Psi(\cdot,\eta)$ satisfies the Dirichlet condition in the
$\xi$-direction at points of $\partial\Omega\setminus\Gamma$.
\vspace{-.8ex}
\end{enumerate}
Because the projection of the set $\Gamma$ onto the
$\eta$-coordinate plane has by assumption zero measure, the
functions $\Psi(\cdot, \eta):\mathbb{R}\to\mathbb{C}$ given by
some $\Psi\in \hat M(\Omega,\Gamma)$ are smooth in $\xi$-direction
for a.e. $\eta\in\hat\omega$ and vanish at
$\partial\Omega\setminus\Gamma$. Hence $Q(\tilde h)$ contains the
subset $\hat M(\Omega,\Gamma)$ which is a core\footnote{The
functions $\Psi\in \hat M(\Omega,\Gamma)$ can have a jump at
$\Gamma$ only in $\eta$-direction.} in the form domain of $\hat
h$. Since $\hat h[\Psi,\Psi]\geq \tilde h[\Psi,\Psi]$ holds for
all $\Psi\in \hat M(\Omega,\Gamma)$ and the norms $|\cdot|_{\hat
h}$ and $|\cdot|_{\tilde h}$ are topologically compatible, it
follows that $Q(\tilde h)\supset Q(\hat h)$. From this we infer
that the inequality $\hat h \ge \tilde h$ is valid. This further
means that the operator
\begin{equation*}
\tilde H = -\Delta^{\mathbb R} \otimes 1_{L^2(\hat\omega)}
- W_-(\xi), \quad \xi \in \mathbb R,
\end{equation*}
associated with $\tilde h$ is strictly bounded by that related to
$\hat h$, i.e.
\begin{equation}\label{EquationHH}
\tilde H < \hat H\,.
\end{equation}
Now we are in position to apply the operator-valued Lieb-Thirring
inequalities \eqref{es12} for $d=1$ and $\sigma \ge 1/2$. In view
of \eqref{EquationHH} and the observed fact that spectra of $H$
and $\hat H$ coincide in the negative part we get
\begin{eqnarray*}
\text{tr } H_-^\sigma &=& \text{tr } \hat H_-^\sigma\\
&\le& \text{tr } \tilde H_-^\sigma \nonumber\\
&\le& r(\sigma,1) L^\mathrm{cl}_{\sigma,1}
\int_{\mathbb R} \mathrm{d}\xi \, \text{tr }W_-^{\sigma+1/2}\nonumber\\
&=& r(\sigma,1) L^\mathrm{cl}_{\sigma,1} \int_{\mathbb R}
\mathrm{d}\xi \, \rm{tr } \left(
-\Delta^{\omega(\xi)}_{\gamma(\xi)} - \lambda_1(\omega_0)
\right)_-^{\sigma+1/2}\,; \nonumber
\end{eqnarray*}
this completes the proof of Theorem \ref{TheoremGeneral}.


\subsection*{Acknowledgments}

The research has been partially supported by Royal Swedish Academy
of Sciences and Academy of Sciences of the Czech Republic within
the exchange program ``Bound states and Resonances in Quantum
Systems and Wave Guides'' and by ASCR within the project K1010104.


\end{document}